\documentclass[reprint,superscriptaddress,showpacs,amsmath,amssymb,aps,pre]{revtex4-1}
\usepackage{graphicx}
\usepackage[justification=raggedright]{caption} 

\usepackage{epstopdf}
\usepackage{amsmath}
\usepackage{diagbox} 
\usepackage{booktabs}
\usepackage{array}         
\usepackage{multirow}     
\usepackage{makecell}     
\usepackage[colorlinks=true, citecolor=blue, urlcolor=blue, linkcolor=blue]{hyperref}
\usepackage{subcaption}
\usepackage{algpseudocode}
\usepackage{xcolor}
\usepackage{algorithmicx}
\usepackage{algpseudocode}
\usepackage{amsmath}
\usepackage[ruled,vlined]{algorithm2e}
\usepackage{float}

\begin{document}

\title{Evolution of cooperation with Q-learning: how much information do we need?}

\author{Yile Ku}
\affiliation{School of Physics and Information Technology, Shaanxi Normal University, Xi'an 710061, P. R. China}
\author{Xin Ou}
\affiliation{School of Physics and Information Technology, Shaanxi Normal University, Xi'an 710061, P. R. China}
\author{Jiqiang Zhang}
\affiliation{School of Physics, Ningxia University, Yinchuan 750021, P. R. China}
\author{Shengfeng Deng}
\affiliation{School of Physics and Information Technology, Shaanxi Normal University, Xi'an 710061, P. R. China}
\author{Huiji Yue}
\email[Email address: ]{yhj2004@snnu.edu.cn}
\affiliation{School of Physics and Information Technology, Shaanxi Normal University, Xi'an 710061, P. R. China}
\author{Li Chen}
\email[Email address: ]{chenl@snnu.edu.cn}
\affiliation{School of Physics and Information Technology, Shaanxi Normal University, Xi'an 710061, P. R. China}

\date{\today}

\begin{abstract}
Cooperation is ubiquitous in both natural and human societies, yet its evolutionary basis remains a major challenge. A long-standing puzzle is whether having more information leads to better decision-making and thus a higher level of cooperation. To address this question, we adopt a recently developed reinforcement learning framework in which individuals learn through trial and error to maximize cumulative rewards -- a paradigm that has successfully explained diverse emergent patterns in human behaviors. Specifically, we equip a structured population with the Q-learning algorithm and systematically vary the size of the interactive neighborhood, which serves as a proxy for perceived information. Interestingly, we observe a non-monotonic relationship between cooperation prevalence and neighborhood size in both two-dimensional square lattices and Barab\'asi–Albert scale-free networks. This inverted U-shaped dependence reveals that an optimal amount of information exists, yielding the highest level of cooperation. Mechanistic analyses show that a moderate neighborhood size enables individuals to strike an optimal balance between information sufficiency and decision-making tractability. This balance allows them to detect reciprocal opportunities while avoiding the deterioration of decision quality due to information overload. Our findings challenge everyday intuition, suggesting that a proper amount of information -- not more -- is optimal for the emergence of cooperation.
\end{abstract}

\maketitle
\section{introduction}\label{sec:introduction}
Cooperative behaviors are pervasive phenomena in nature and human societies~\cite{maynard1997major,cheney2011extent,dawkins2006selfish}. In nature, the division of labor within bee colonies ensures colony survival, and food sharing in vampire bats helps individuals endure hunger crises~\cite{dugatkin1997cooperation,wilkinson1984reciprocal}. In human societies, cooperation profoundly shapes individual behavior and social interactions~\cite{zelenski2015cooperation}. However, its prevalence challenges the core logic of Darwinian evolution~\cite{dawkins2006selfish}. Darwinism emphasizes that individuals adapt to their environment by maximizing their own survival and reproductive success, whereas cooperation is inherently altruistic -- individuals provide benefits to others at a personal cost~\cite{fehr425nature}. This places cooperators at a competitive disadvantage against selfish individuals, making cooperation difficult to evolve. Consequently, understanding the emergence and maintenance of cooperation remains a pressing challenge~\cite{Pennisi2006}.

Evolutionary game theory provides a fundamental framework for addressing this puzzle by integrating the analytical tools of classical game theory into evolutionary dynamics~\cite{Smith1982,Taylor1978}. Within this framework, early studies on prototypical game models, such as the prisoner's dilemma~\cite{Poundstone1993,Doebeli2005,Axelrod1980a,Perc2008,Axelrod1980b,Milinski1998}, have revealed several key mechanisms for the evolution of cooperation~\cite{nowak2006five}, including direct reciprocity~\cite{Trivers1971,Pacheco2008}, indirect reciprocity~\cite{Nowak1998,Ohtsuki2005,Brandt2006,Ghang2015}, network reciprocity~\cite{Nowak1992a}, group selection~\cite{Keller1999,Queller1992}, and kin selection~\cite{Hamilton1963,Hamilton1964a,Hamilton1964b,Griffin2002}. Subsequent research has further uncovered additional mechanisms, such as reward~\cite{Sigmund2001}, punishment~\cite{Yang2018,Takesue2018,Szolnoki2013}, reputation~\cite{Fu2008,Xia2023}, and dynamical reciprocity~\cite{liang2022dynamical}, \emph{etc.}.

Despite this progress, one long-standing question remains open in the community: Does more information necessarily lead to higher levels of cooperation? Intuitively, having more information provides individuals with a more complete picture of their surroundings, which would be expected to enable wiser decision-making and potentially promote cooperative behaviors. However, an early study~\cite{ifti2004effects} suggests that this may not hold true: when the neighborhood size -- interpreted as a proxy for information -- exceeds a critical value, the population always converges to the mean-field limit of no cooperation on lattice structures. Subsequent studies~\cite{Szabo2009,Zhu2013,Qian2015,Wang2019} quantitatively corroborated this finding and further revealed that an intermediate neighborhood size is optimal for fostering cooperation. The key mechanism underlying the breakdown of cooperation under large neighborhoods lies in the failure of network reciprocity~\cite{Nowak1992a}: cooperation thrives when cooperators form compact clusters, but large neighborhoods disrupt this compactness, thereby undermining cooperation. To achieve an appropriate neighborhood size, adaptive adjustment strategies have often proven effective~\cite{Han2021,Lu2023,Ma2024}.
It is important to note, however, that all these studies were conducted within the imitation learning (IL) paradigm, in which players copy successful neighbors according to fixed update rules. A notable limitation of this paradigm is that many behavioral experiments exhibit strategic changes that cannot be explained by imitation alone~\cite{traulsen2010human}, and network reciprocity often fails to account for the observed patterns~\cite{gracia2012heterogeneous,sanchez2018physics}. Consequently, the role of information quantity in the evolution of cooperation remains an unresolved issue.

Recently, reinforcement learning (RL)~\cite{Sutton2018, zheng2026brief} has emerged as a distinct paradigm for understanding evolutionary game dynamics. In the RL framework, players make decisions by learning through trial and error to maximize accumulated rewards over the long term. Unlike the ``follow-the-peer" social heuristic characteristic of IL, RL players develop their own decision-making strategies through interaction with the environment -- an introspective learning process that is intrinsic to humans and many other species. This fundamental difference sets the RL paradigm apart from the IL paradigm and makes it particularly well-suited for modeling human behavior, as its foundational principles are supported by neuroscientific evidence~\cite{Lee2012, Rangel2008}. Indeed, a growing body of work has revisited classic emergence problems within evolutionary games using RL, including the emergence of cooperation~\cite{Zhang2020,Jia2021,Wang2023,Song2022,He2022,Wang2023,Ding2023,Geng2022,zheng2024evolution,Yang2024,Zhang2024}, fairness~\cite{Zheng2025fairness}, trust~\cite{Zheng2024Trust}, resource allocation~\cite{Andrecut2001,Zhang2019,Zheng2023resource,Zhang2024b}, punishment~\cite{Zhao2024punishment}, and even biodiversity in ecology~\cite{jiang2026decoding}. These studies suggest that RL may offer a unified framework for addressing a wide range of societal puzzles. Returning to our central question: within the RL paradigm, \emph{does more information necessarily lead to higher levels of cooperation}? Or put the way around, how much information do we need to make proper decisions that sustain a satisfactory level of cooperation?

In this work, we employ Q-learning~\cite{Watkins1992, Watkins1989} -- a classic RL algorithm -- to investigate the role of information quantity in the evolution of cooperation using prisoner's dilemma game. Specifically, we consider a two-dimensional square lattice in which the information acquisition range is controlled by varying the neighborhood size, with the number of neighbors ranging from 1 to 12. Interestingly, we uncover the existence of an optimal information quantity that maximizes the cooperation level, occurring at around $3\sim4$ neighbors. This finding remains robust when the underlying topology is replaced by Barab\'asi–Albert scale-free networks. We further elucidate the mechanisms behind this phenomenon by analyzing the dynamics of the evolution and a semi-analytic theory.

\section{model} \label{sec:model}
In our work, the population is placed on a two-dimensional square lattice of size \(N = L \times L\) with periodic boundary conditions. 
For each pair of players, they engage in the prisoner's dilemma game (PDG), in which each player can either cooperate (C) or defect (D). Mutual cooperation yields a reward $R$, mutual defection brings punishment $P$, and if one cooperates while the other defects, the defector receives the temptation payoff $T$, whereas the cooperator receives the sucker's payoff $S$. In the strong version of PDG, the four payoffs satisfy the ordering \(T > R > P > S\) and \(S + T < 2R\), which creates a fundamental dilemma: although cooperation benefits the collective, defection remains the individually superior choice.
In practice, the payoff matrix is often parameterized as follows:
\begin{equation}
	\Pi = \begin{pmatrix}
		\Pi_{CC} & \Pi_{CD} \\
		\Pi_{DC} & \Pi_{DD}
	\end{pmatrix} = \begin{pmatrix}
		R & S \\
		T & P
	\end{pmatrix}= \begin{pmatrix}
		1 & -b \\
		1+b & 0
	\end{pmatrix},
\label{eq:payoff_matrix}
\end{equation}
where \(b \in (0,1)\) controls the intensity of the dilemma; a larger value of \(b\) corresponds to a weaker inclination toward cooperation. 

In our study, players employ the Q-learning algorithm with an $\epsilon-$greedy policy to guide their decision-making. Specifically, each individual maintains a Q-table (Table \ref{qtablewithnei}), which is a 2d matrix of Q-values spanned by the state and action sets. Here, the action set is \(\mathcal{A} = \{\text{C}, \text{D}\}\), while the state set \(\mathcal{S} = \{0, 1, 2, \dots, k\}\) is defined by the number of cooperators among the focal individual's neighbors, with $k$ denoting the neighorhood size. 
The value \(Q_{s,a}\) is the state-action value function, which evaluates the value of taking action \(a\) in state \(s\); a higher value for a given action \(a\) in a particular state \(s\) indicates that it is more likely to be selected over the alternative. 

\begin{figure}[tbp]
\centering
\includegraphics[width=0.33\textwidth]{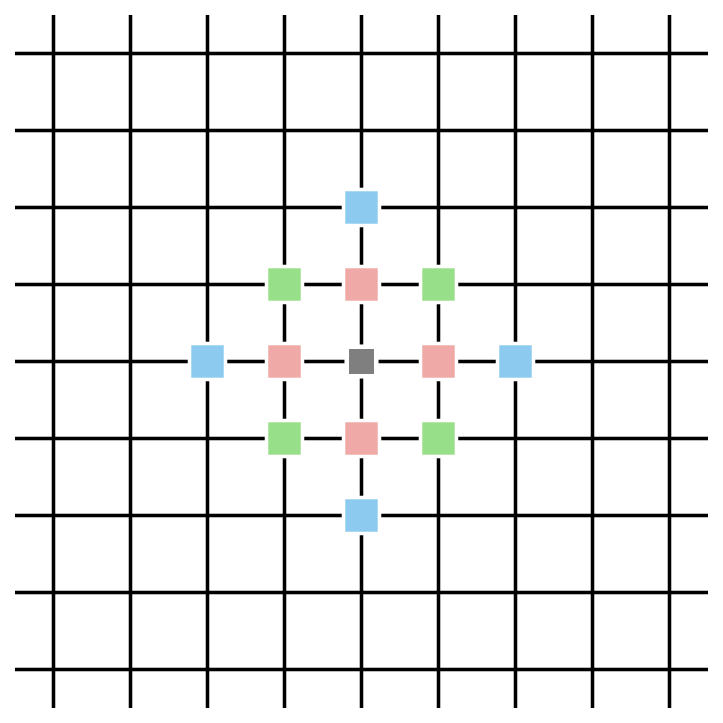}
\caption{Scheme of the neighborhood structure on a two-dimensional lattice. 
For \(k = 1\sim4\), neighbors are randomly drawn from the four nearest neighbors (pink). 
For \(k = 5\sim8\), the four nearest neighbors are included, and the remaining neighbors are randomly selected from the four diagonal nearest neighbors (green). 
For \(k = 9\sim12\), all eight nearest and diagonal nearest neighbors are included, and the remaining neighbors are randomly chosen from the next-nearest neighbors (blue). 
}
\label{fig:neighborhood}
\end{figure}

The evolutionary process follows a synchronous update scheme. Initially, each player is assigned a random strategy from the set \(\mathcal{A}\), and all values \(Q_{s,a}\) in the Q-table are randomly initialized between 0 and 1, reflecting essentially no initial preference for any particular action.
To investigate the impact of information quantity,  the neighborhood size $k$ is varied from 1 to a maximum of 12 neighbors. For a given $k$, each player randomly selects its neighbors but sequentially from near to far, as illustrated in Figure \ref{fig:neighborhood}.

In each round $t$, the update proceeds as follows: (i) each individual chooses a random action \(a_t\in\mathcal{A}\) with probability \(\varepsilon\); otherwise, it follows the guidance of its Q-table by selecting the action $a$ with the higher Q-value within the current state $s_t$, i.e. \(Q_{s,a} > Q_{s,\hat{a}}\). 
(ii) After all individuals have made their decisions, they transition to their new state $s_{t+1}$, and their average payoffs $\bar{\pi}$ are computed as 
\begin{equation}
	\bar{\pi}_i = \frac{1}{k_i} \sum_{j \in \Omega_i} \Pi(a_i, a_j),
\end{equation}
where \(k_i\) is the number of neighbors of player \(i\), \(\Omega_i\) denotes its neighbor set, and $\Pi$ is the payoff matrix defined in Eq. (\ref{eq:payoff_matrix}).
(iii) They draw lessons by revising the corresponding Q-values using the Bellman equation:
\begin{equation}
	Q(s_t, a_t) \leftarrow (1 - \alpha) Q(s_t, a_t) + \alpha \left[ \bar{\pi}_t + \gamma \max_{a'} Q(s_{t+1}, a') \right],
\end{equation}
where \(s_t\) and \(a_t\) are the current state and action, respectively, and \(s_{t+1}\) is the resulting new state. 
The parameter \(\alpha \in (0, 1]\) is the learning rate, capturing the contribution of the current experience; a smaller $\alpha$ indicates the past experience is replaced more slowly. The discount factor \(\gamma \in [0, 1)\) determines the weight assigned to future rewards, with \(\max_{a'} Q(s_{t+1}, a')\) representing the expected maximal reward in the next round.

The evolution, consisting of steps (i)-(iii), is repeated until the system reaches a steady state or the predefined maximum number of rounds is attained.
Unless otherwise stated, the learning parameters are fixed at typical values \(\alpha = 0.1\), \(\gamma = 0.9\), \(\varepsilon = 0.01\), with \(b = 0.1\) and \(L = 50\). In our simulations, the evolution runs for $2\times10^{6}$ rounds, and we sample data over the final $1\times10^{4}$ rounds.

In parallel, we also examine Barab\'asi-Albert (BA) scale-free networks~\cite{Barabasi1999Emergence} to assess the generality of our findings. Compared to the two-dimensional lattice, BA networks are characterized by a heterogeneous degree distribution and are widely regarded as a representative model for many real-world systems. Due to this inherent degree heterogeneity, individuals in BA networks naturally possess varying numbers of neighbors, implying that different nodes may acquire substantially different amounts of information. In particular, hub nodes have much larger neighborhoods than peripheral nodes, thus obtaining more information and maintaining larger Q-tables. This naturally raises a question: are hub nodes necessarily more cooperative since they possess more information?

\begin{table}[t]
	\centering
	\small  
	\renewcommand{\arraystretch}{1.2}  
	\begin{tabular}{p{2.8cm}p{2.3cm}p{2.3cm}}
		\hline\hline
		\multirow{2}{*}{\textbf{State}} & \multicolumn{2}{c}{\textbf{Action}} \\
		& \textbf{$a_1 = C$} & \textbf{$a_2 = D$} \\
		\hline
		\( s_0( n_{c} = 0) \) & \( Q(s_0, C) \) & \( Q(s_0, D) \) \\
		\( s_1(n_{c} = 1) \) & \( Q(s_1, C) \) & \( Q(s_1, D) \) \\
		\(\vdots\) & \(\vdots\) & \(\vdots\) \\
		\( s_{k}(n_{c} = k) \) & \( Q(s_{K}, C) \) & \( Q(s_{K}, D) \) \\
		\hline\hline
	\end{tabular}
	\captionsetup{position=bottom}
	\caption{Q-table for each player. The states \(s_j ~(j={0,\dots,k})\) correspond to the number of cooperators $n_c$ in its neighborhood, and \(k\) is the neighborhood size. There are two actions available: cooperation (C) and defection (D).}
	\label{qtablewithnei}
\end{table}

\section{results}\label{sec:result}

\begin{figure}[htbp]
\centering
\includegraphics[width=\columnwidth]{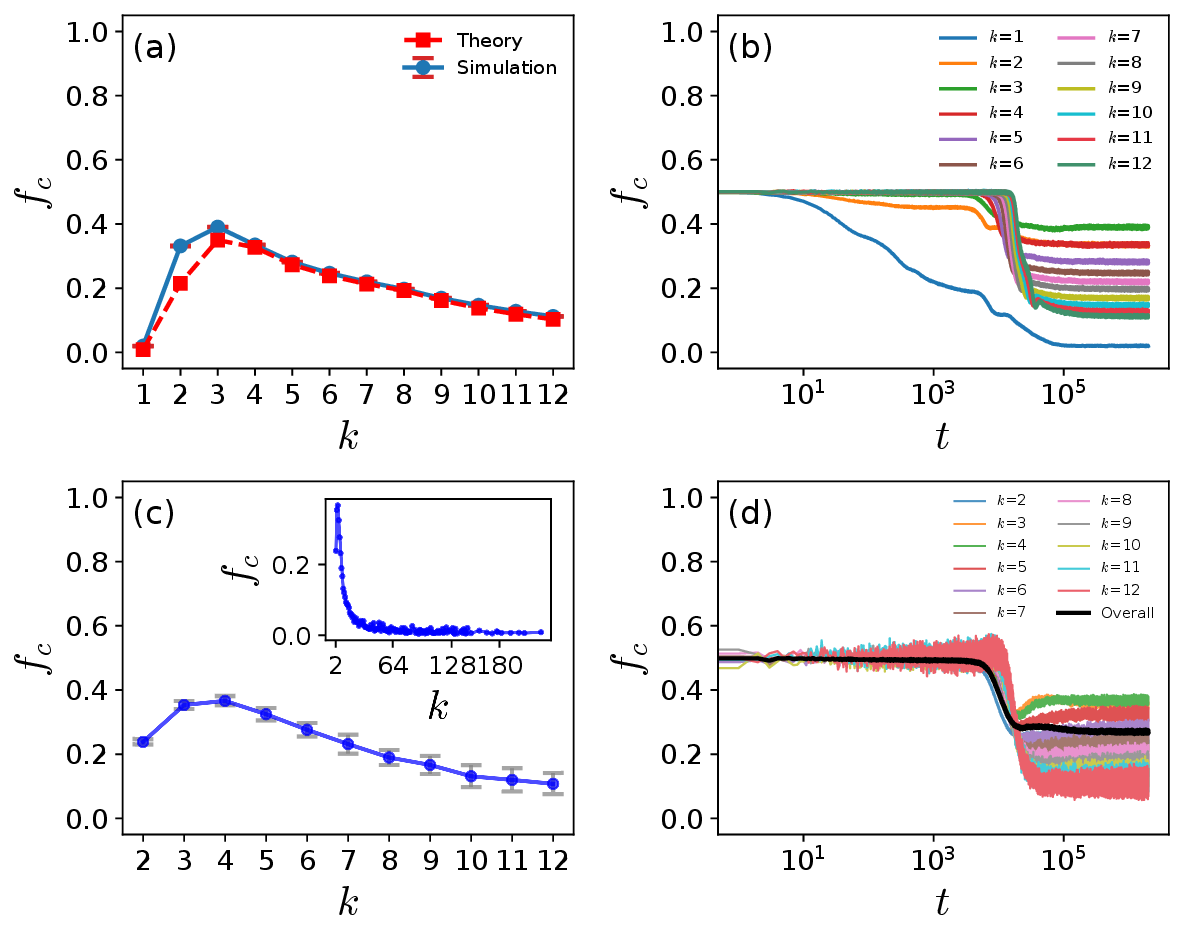}
\caption{Impact of information on cooperation.
(a,c) shows the dependence of the cooperation level on information quantity for 2d lattice and BA networks, respectively. 
In (a), the control parameter $k$ is the neighbourhood size varied in our study, \(f_{C} = N_{C} / N\), where \(N_{C}\) is the number of cooperators and \(N = L \times L\) is the total population size.
For BA networks, $k$ represents the node degree, and \( f_{C}(k) = N_{C}(k) / N_k \), where \( N_k \) is the total number of individuals with degree \(k\) and \(N_C(k)\) is the number of cooperators among them.
The red dashed line in (a) is our semi-analytical treatment (Sec.~\ref{sec:theory}).
 Each data point is averaged over the last \(1\times10^{4}\) time steps and over 50 independent realizations; error bars indicate the standard deviations. The inset in (c) shows the dependence over a broader range of degrees in BA networks. 
(b,d) present the typical time series of the cooperation level \(f_{C}\) for different information quantities, with each data point averaged over 50 ensembles. 
Other parameters: \(b = 0.1\), \(\alpha = 0.1\), \(\gamma = 0.9\), \(\varepsilon = 0.01\), \(t_{\text{max}} = 2\times10^{6}\).
}
\label{fig:coop_time_series}
\end{figure}

We first report the dependence of cooperation level \(f_{C}\) on the amount of information in the 2d lattice population, where the information quantity is modulated by the neighborhood size $k$, and \(f_{C}\) is measured across the entire population.
As shown in Fig.~\ref{fig:coop_time_series}(a), the cooperation level increases with 
$k$ when information is highly limited. However, an optimal neighborhood size emerges around 
$k=3$, at which $f_C$ reaches its maximum; further enlarging the neighborhood instead reduces cooperation. This trend is corroborated by the typical time series in Fig.~\ref{fig:coop_time_series}(b), where 
$f_C$  for $k=3$ remains the highest after stabilization. These results suggest that cooperation is hindered by both insufficient and excessive information, and that an intermediate amount of information maximizes the cooperative level in the population.

\begin{figure*}[ht]
\centering
\includegraphics[width=2\columnwidth]{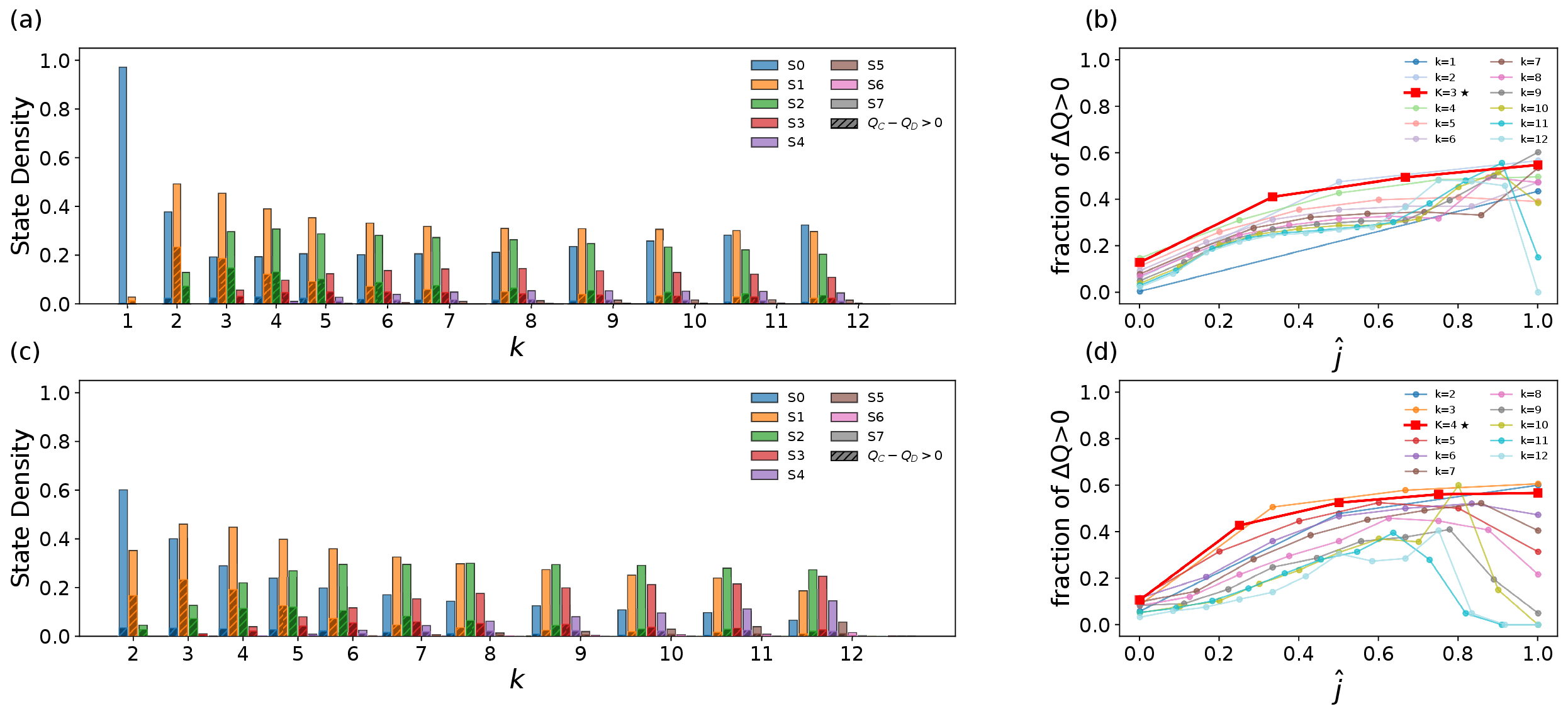}
\caption{Cooperation preference analysis. 
(a) The state densities for different neighborhood sizes $k$ on the 2d square lattice; the shaded portion of each bar represents the cooperation preference, defined as the fraction of cases with ${\Delta Q}_{s_j} > 0$. 
(b) The cooperation preference on 2d lattice, i.e., the fraction of ${\Delta Q}_{s_j} > 0$ within each state, where the state labeling is normalized by \(\hat{j}=j/k\). 
(c, d) show the corresponding results for BA networks. 
All data are averaged over the last $1\times10^{4}$ steps after reaching steady state, and over 50 independent realizations. 
Other parameters: $b = 0.1$, $\alpha = 0.1$, $\gamma = 0.9$, $\varepsilon = 0.01$.
}
\label{fig:coop_willingness}
\end{figure*}

To examine the robustness of this finding, we extend our analysis to BA scale-free networks, where the degree 
$k$ serves as the control parameter for information availability. Note that in this case, the cooperation level is defined at the subpopulation level: \( f_{C}(k) = N_{C}(k) / N_k \), with \(N_C(k)\) denoting the number of cooperators among nodes of degree $k$, and \( N_k \) the total number of such nodes. Fig.~\ref{fig:coop_time_series}(c) demonstrates that the optimal information effect persists in BA networks, although the peak shifts to $k=4$. The corresponding time series in Fig.~\ref{fig:coop_time_series}(d) further corroborates this observation, and the average cooperation level across the entire population is also displayed.

A particularly intriguing finding in the BA network is illustrated in Figs.~\ref{fig:coop_time_series}(c) and~\ref{fig:coop_time_series}(d): low-degree (peripheral) nodes sustain relatively high cooperation levels, whereas high-degree (hub) nodes exhibit cooperation levels near zero, as shown in the inset of Fig.~\ref{fig:coop_time_series}(c). This pattern stands in sharp contrast to the established result with the imitation learning~\cite {Santos2005}, where hubs typically maintain high levels of cooperation and serve as the backbone of cooperative behaviors within the whole population. Such reversed cooperation prevalence between hubs and periphery nodes highlights a fundamental distinction in the underlying mechanisms governing the evolution of cooperation between IL and RL paradigms~\cite{zheng2026brief}.

\section{mechanism analysis}\label{sec:mechanism}

To understand the dynamical mechanisms underlying the above phenomena, we focus on the evolution of the Q-table and compute the Q-value difference within each state as follows
\begin{equation}
        \Delta Q_{s_j} = \frac{1}{N} \sum_{i=1}^{N} \left( Q_{s_j,C}^i - Q_{s_j,D}^i \right).
\end{equation}
According to the working logic of Q-learning, \(\Delta Q_{s_j} > 0\) implies that individuals in state \(s_j\) are more inclined to choose cooperation on average; conversely, \(\Delta Q_{s_j} < 0\) indicates defection is preferred. Thus, this Q-value difference can be interpreted as a measure of \emph{cooperation preference}.

We present the state density for all states and the fraction of individuals with a positive cooperation preference of \(\Delta Q_{s_j} > 0\) across all neighborhood sizes on the 2d square lattice in Fig.~\ref{fig:coop_willingness}(a). A separate version for each $k$ is provided in Fig.~\ref{fig:lattice_pdf} in Appendix~\ref{sec:Appendix_pdf}. We observe that, in most cases, the state densities are dominated by \(s_{0,1,2,3}\), while the densities for highly cooperative states \(s_{4,5,...}\) are nearly zero. The positive cooperation preference is coded by the shaded region within each bar, though this visualization makes it inconvenient to directly compare preferences for different states and neighborhood sizes.

For a more systematic comparison, we normalize the state index by the neighborhood size, defining  \(\hat{j}=j/k\), and examine the cooperation preference for states \(s_{\hat{j}}\). This normalization allows us to compare cooperation preferences (i.e., the fraction of individuals with  \(\Delta Q_{s_j} > 0\)) across different information scenarios with varying $k$. In this way, Fig.~\ref{fig:coop_willingness}(b) reveals two prominent properties.
(i) The cooperation preference monotonically strengthens with increasing normalized state index \(\hat{j}\), meaning that players tend to defect in predominantly defective surroundings (small \(\hat{j}\)) but cooperate in cooperative environments (large \(\hat{j}\)). This observation aligns with previous theoretical results~\cite{Sheng2024Catalytic,Zhao2024Emergence,Zhao2025evolution} and is consistent with conditional cooperation behavior documented in behavioral economics~\cite{Fischbacher2001Are}. (ii) More importantly, individuals with neighborhood size \(k=3\) exhibit the highest cooperative preference compared to other sizes across almost all states. Although this advantage appears only slight in each state, the cumulative effect across all states ultimately leads to a substantially higher overall level of cooperation. It is also worth noting that, for \(\hat{j}=1\), the cooperation preference is strong (the fraction of $\Delta Q_{s_j} > 0$ is generally the highest); however, the densities of these states are very small, especially for large $k$, rendering their practical impact negligible, see e.g. Figs.~\ref{fig:coop_willingness}(a,c).

Two ensemble-averaged time evolution of \(\overline{\Delta Q}_{s_j}\) for each state, corresponding to \(k=3\) and \(4\), are provided in Fig.~\ref{fig:deltaQ} (Appendix~\ref{sec:Appendix_deltaQ}). They show that the average cooperation preference across all states is low, as \(\overline{\Delta Q}_{s_j}<0\) for all states, which explains why cooperation levels are modest even under these two moderate information conditions. Comparatively, the more defective the neighborhood (e.g., $s_{0,1}$), the more negative the value of \(\overline{\Delta Q}_{s_j}\). 
It should be emphasized that \(\overline{\Delta Q}_{s_j}\) represents the mean over all players; even when \(\overline{\Delta Q}_{s_j} < 0\), some individuals still exhibit cooperative preferences ($\Delta Q_{s_j} > 0$), as shown in Fig.~\ref{fig:coop_willingness}.

We apply the same analysis to the BA networked population, and the results are presented in Figs.~\ref{fig:coop_willingness}(c,d). Specifically, the two key properties i) and ii) largely hold, except for a noticeable decay at large \(\hat{j}\), particularly for large $k$ in Fig.~\ref{fig:coop_willingness}(d). This decline stems from the fact that high-degree nodes are rare in BA networks due to the power-law degree distribution, and even rarer are those whose neighborhoods are highly cooperative (i.e., \(\hat{j}\rightarrow 1\)). This scarcity is more clearly observed in Fig.~\ref{fig:BA_pdf} (Appendix~\ref{sec:Appendix_pdf}), where the densities for states $s_{5,6,...}$ become vanishingly small. Another observation is that, although the cooperation preference for $k=3$ remains slightly higher than that for $k=4$ in some states, the state density distribution for $k=4$ is more concentrated in states with higher cooperation preference, ultimately yielding a slightly higher overall cooperation level.

\begin{figure*}[htbp]
\centering
\includegraphics[width=1.0\textwidth]{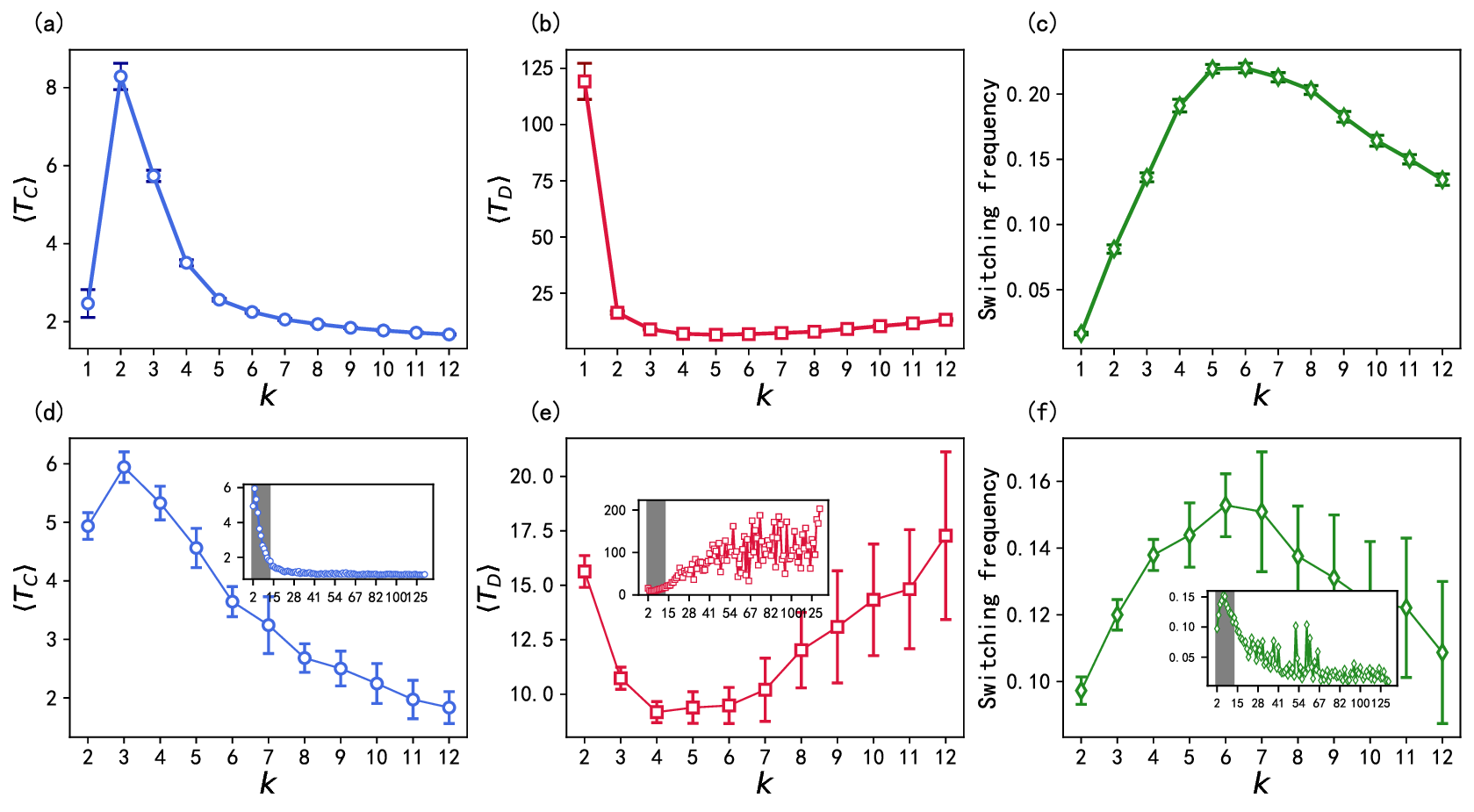}
\caption{Strategy persistence and switching.
(a, d) Average cooperation duration $\langle T_C\rangle$; (b, e) Average defection duration $\langle T_D\rangle$; (c, f) Strategy switching frequency. 
The upper row (a-c) corresponds to the 2d square lattice, and the bottom row (d-f) to the BA network. The insets in (d-f) show the full range of degrees, with the shaded regions indicating the enlarged portions displayed in the main panels.
All data are averaged over the last \(1\times10^{4}\) time steps after reaching the steady state at \(t_{\text{max}} = 2\times10^{6}\), and over 50 independent realizations. Error bars denote the standard deviation of the mean across 50 realizations. 
Other parameters: \(b = 0.1\), \(\alpha = 0.1\), \(\gamma = 0.9\), \(\varepsilon = 0.01\).
}
\label{fig:strategy_persistence}
\end{figure*}

To further understand why a moderate amount of information yields optimal cooperation, we compute the average cooperation duration $T_C$ and defection duration $T_D$ -- defined as the number of consecutive steps an individual remains in cooperation or defection, respectively, and the strategy switching frequency, i.e., the number of strategy switches divided by the total number of evolution steps. Their dependence on the neighborhood size $k$ is shown in Fig.~\ref{fig:strategy_persistence} for both the lattice and BA networks.

When information is highly limited, e.g., the case of $k=1$ on the two-dimensional lattice, Figs.~\ref{fig:strategy_persistence}(a--c) show that defection dominates, with 
$T_D\gg T_C$, and the corresponding switching frequency is around 0.02, indicating that cooperation arises solely from exploration at rate 
\(\varepsilon = 0.01\). This observation is consistent with Fig.~\ref{fig:lattice_pdf}(a) (Appendix~\ref{sec:Appendix_pdf}), confirming that individuals persistently defect in this information-limited scenario.

When the information increases to $k=2$, $T_C$ rises to about 8 steps, $T_D$ drops significantly to approximately 20 steps, and the strategy switching frequency becomes about 0.089. This indicates that cooperation begins to emerge and stabilize, and individuals exhibit relatively persistent strategy choices -- whether cooperating or defecting -- as switching remains infrequent.

Further increasing $k$ reduces both $T_C$ and $T_D$, accompanied by an increase in switching frequency, implying that strategy persistence weakens and individuals switch more frequently between cooperation and defection. This trend, however, reverses for $k>6$, where $T_D$ increases again and switching frequency declines. In the extreme case of 
$k=12$, for instance, individuals remain cooperative for only one step before reverting to defection. Yet the switching frequency is much higher than the exploration rate ($0.14\gg 0.01$), indicating that individuals do not settle into stable defection but instead switch strategies repeatedly. Thus, in this information-overloaded scenario, a stable policy fails to form, and players behave chaotically with frequent strategy changes.

For the moderate case (here $k=3$), the cooperation duration is $T_C\approx6$, and the defection duration is $T_D\approx15$ steps. On the one hand, this means individuals have developed clearer preferences compared to the information-overloaded scenario, as they maintain either cooperation or defection for several consecutive steps. On the other hand, they are not as rigid in their strategies as those in the information-limited scenarios ($k=1,2$); instead, they adjust their strategies to better adapt to the changing environment.

Figs.~\ref{fig:strategy_persistence}(d--f) show that the above evolutionary properties are qualitatively similar in BA networks. Specifically, Fig.~\ref{fig:strategy_persistence}(d) shows a peak in cooperation duration $T_C$ at $k=3$, after which it decays with increasing degree and approaches zero for high-degree nodes, whereas Fig.~\ref{fig:strategy_persistence}(e) exhibits the opposite trend for $T_D$, with a minimum at $k=4$. The insets in Figs.~\ref{fig:strategy_persistence}(d) and (e) clearly reveal that hubs are persistent defectors, with $T_C\rightarrow 0$ and $T_D$ becoming very large for nodes with large degree $k$.

These two profiles of $T_C$ and $T_D$ naturally explain why an intermediate degree yields the highest level of cooperation, for the same reasons discussed above. In information-limited scenarios (e.g., $k=2$), individuals strongly prefer defection with high persistence. In information-overloaded cases, policies are not well-formed: although defection is still preferred ($T_D\gg T_C$), persistence is lower due to more frequent switching. From a learning perspective, the state space becomes too large in such scenarios, and within a reasonably long evolution, learning remains incomplete, resulting in near-random behavior.


In brief, optimal cooperation under moderate information arises from a balance between information sufficiency and decision-making tractability. Insufficient information is cognitively manageable but biased toward defection, whereas excessive information exceeds individuals' learning capacity, preventing the formation of effective policies and thereby also leading to poor cooperative outcomes.

\section{A semi-analytic theory}\label{sec:theory}

To further explore the underlying mechanism quantitatively, we implement a semi-analytical treatment for the case of the two-dimensional square lattice. Within the framework of a Markov process, we construct the state transition matrix to study the evolution of different state densities. By incorporating the cooperation preference information extracted from simulations, we can then solve for the stationary densities of different states, as well as the overall cooperation level $f_C$. Detailed derivations are provided in Appendix~\ref{sec:Appendix_theory}. From a physics perspective, this treatment neglects the underlying spatial structure of the population and can thus be viewed as a mean-field approximation.

\begin{figure}[tbp]
\centering
\includegraphics[width=0.49\textwidth]{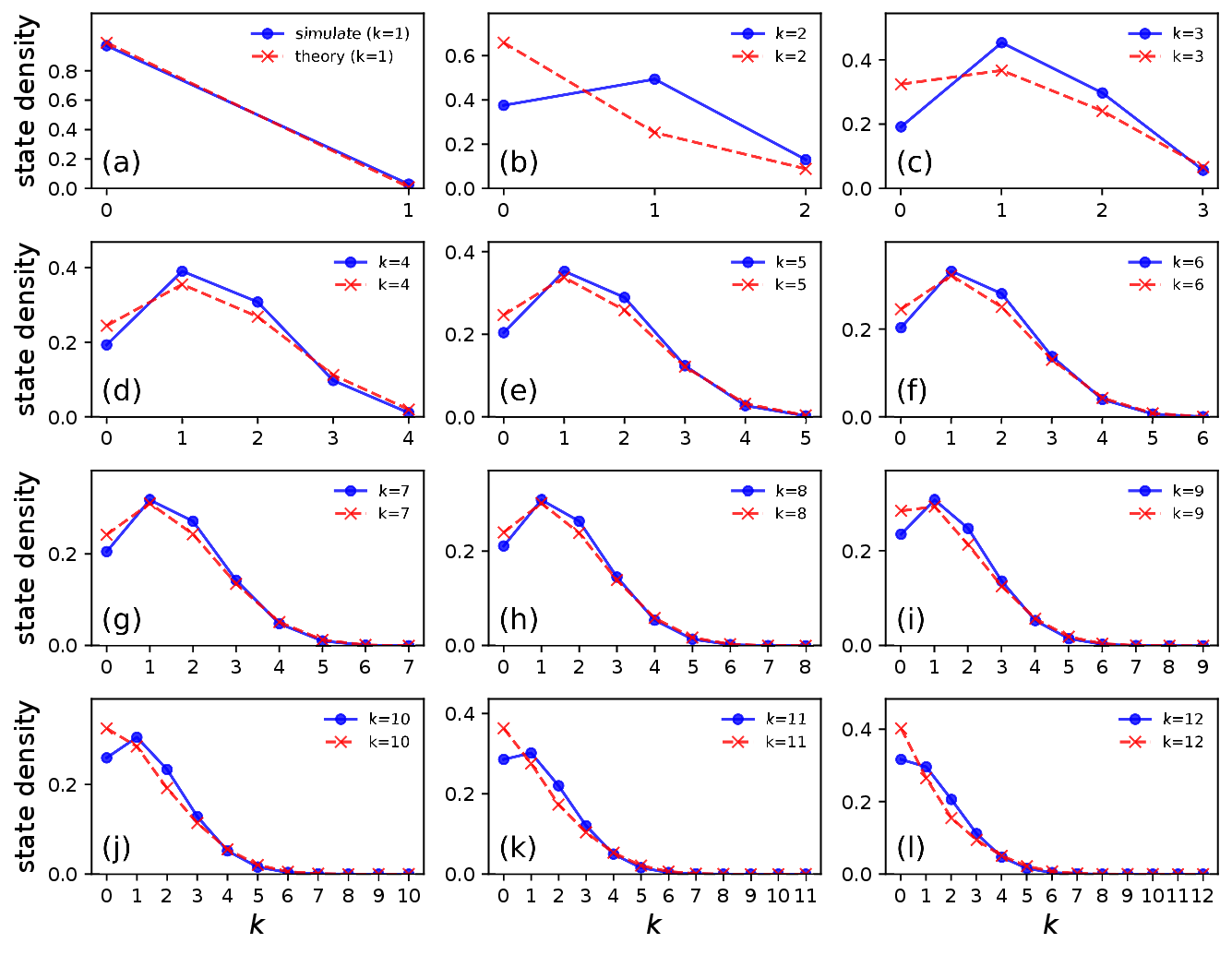}
\caption{State densities for different numbers of neighbors on 2d lattice. Subplots (a-l) are for neighborhood sizes $k=1\sim12$; both simulation and theoretical results are shown. All the setups are the same as Fig.~\ref{fig:coop_time_series}(a)}
\label{fig:sim_th_state}
\end{figure}

The results of our theory are illustrated in Fig.~\ref{fig:coop_time_series}(a), which correctly captures the overall trend of $f_C$ as a function of neighborhood size $k$ observed in simulations. In particular, it accurately reproduces the optimal cooperation level at the moderate $k$ values. Since the main logic of our theory is based on the computation of state densities, the theoretical stationary-state distributions is also presented in Fig.~\ref{fig:sim_th_state} and compared with the simulations. Although modest discrepancies exist between the theoretical and numerical outcomes, their overall evolutionary tendencies remain consistent. Specifically, the theoretical results are in perfect agreement with the numerical results for the extreme case of $k=1$ [Fig.~\ref{fig:sim_th_state}(a)], as no spatial correlation is present. However, this agreement does not hold well for intermediate cases, such as $k=2,3,4$, where strategic correlations among neighboring agents become prominent and violate the independence assumption underlying the Markovian process, leading to larger deviations [Figs.~\ref{fig:sim_th_state}(b--d)]. As $k$ is further increased [Figs.~\ref{fig:sim_th_state}(e--l)], the population approaches the well-mixed assumption, and the deviations between theoretical and simulation results shrink accordingly.

\section{Extension}\label{sec:extension}

\begin{figure}[tbp]
\centering
\includegraphics[width=\columnwidth]{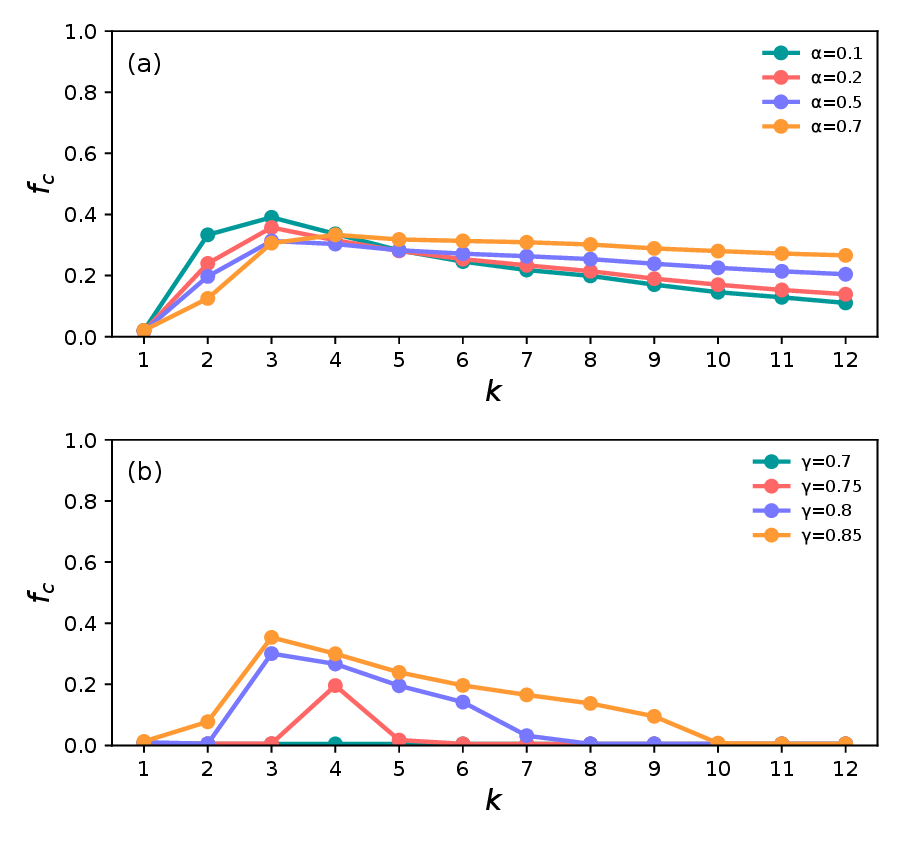}
\caption{The impact of learning parameters for the lattice case. 
(a) The dependence of $f_C$ on the neighorhood size $k$ for different \(\alpha\) by fixing \(\gamma = 0.9\). 
(b) The dependence of $f_C$ on $k$ for different \(\gamma\) by fixing \(\alpha = 0.1\). 
All data are averaged over the last \(1\times10^{4}\) time steps after reaching state state at \(t_{\text{max}} = 2\times10^{6}\), and over 10 independent realizations. 
Other parameters: \(\varepsilon = 0.01\), \(b = 0.1\).
}
\label{fig:a_gamma}
\end{figure}

To examine the impact of the learning parameters in Q-learning, we conduct additional simulations on the two-dimensional lattice using various combinations of parameter values. Fig.~\ref{fig:a_gamma}(a) shows that increasing the learning rate $\alpha$ reduces the maximal cooperation level $f_C$ for small neighborhoods (small $k$), but enhances cooperation for larger neighborhoods (large 
$k$). These contrasting effects arise from the nuanced role of $\alpha$ in the learning process: a small $\alpha$ helps preserve past experience but slows down learning, thereby requiring longer training durations; in contrast, a larger $\alpha$ discards experience more rapidly but accelerates learning, which often proves advantageous when learning samples are limited. In small-neighborhood scenarios, learning samples are abundant, so increasing $\alpha$ impairs cooperation. However, for large $k$, where many states suffer from insufficient samples, a higher $\alpha$ facilitates faster learning and thus promotes cooperation.

In contrast, the discount factor $\gamma$ exerts a more pronounced influence on the cooperation level, as shown in Fig.~\ref{fig:a_gamma}(b). Reducing $\gamma$ suppresses overall cooperation, and when 
$\gamma\le 0.7$, cooperation becomes negligible for all values of $k$. This occurs because the emergence of cooperation in RL relies on players maintaining a long-term perspective; when $\gamma$ decreases, individuals become overly focused on immediate rewards, undermining the viability of cooperative strategies.

Overall, while both learning parameters modulate the absolute level of cooperation, they do not alter the qualitative dependence of $f_C$ on the neighborhood size $k$. The presence of optimal neighborhood size that maximizes cooperation remains robust across the parameter space examined.

\section{Conclusion}\label{sec:conclusion}

This study introduces the reinforcement learning (RL) paradigm into the prisoner's dilemma game to systematically examine how information quantity -- operationalized as neighborhood size -- affects the evolution of cooperation. Unlike imitation learning, where individuals copy successful peers, Q-learning agents make decisions based on accumulated payoffs through trial and error. Through simulations on both 2d regular lattices and Barab\'asi--Albert scale-free networks, we uncover a non-monotonic, inverted U-shaped relationship between cooperation level and information availability, with cooperation maximized at an intermediate neighborhood size around $3\sim4$.

Our mechanistic analysis reveals that optimal cooperation emerges when individuals strike a balance between information sufficiency and decision-making tractability. Strategy persistence analysis further shows that, under moderate neighborhood sizes ($k=3,4$), a subset of individuals can sustain cooperation with relatively stable strategies while retaining necessary strategic flexibility. In contrast, under either insufficient or excessive information, strategies become either rigidly fixated on defection or excessively volatile -- both of which prevent the emergence of stable cooperation. We develop a semi-analytic theory that successfully reproduces this non-monotonic dependence.

Importantly, although previous studies~\cite{ifti2004effects,Szabo2009,Zhu2013,Qian2015,Wang2019} reached similar conclusions, the underlying mechanisms differ fundamentally, especially in the information-overloaded scenario. In those works, the breakdown of cooperation is attributed to the failure of network reciprocity stemming from the disrupted compactness of cooperative clusters. In contrast, the emergence and maintenance of cooperation in our RL-based framework arise from information sufficiency and cognitive tractability -- no obvious spatial patterns are required or observed.

To further validate the findings of this study, controlled behavioral experiments are called for. By manipulating the information acquisition range, such as the size of the interactive neighborhood, in experimental settings, the dependence of cooperation prevalence on information quantity can be directly examined. In brief, our study suggests that making proper decisions does not necessarily require more information -- a reassuring message in today's era of information overload.

\bibliographystyle{unsrt}
\bibliography{references}

\section*{Data and code availability}
All data in our study are generated by our code written in Python, which is available at \href{https://github.com/chenli-lab/RL-information}{https://github.com/chenli-lab/RL-information}.

\section*{ACKNOWLEDGEMENTS}
This work is supported by the National Natural Science Foundation of China (Grants Nos. 12075144, 12165014), 
the Fundamental Research Funds for the Central Universities (Grant Nos. GK202401002, GK202406016), and the Natural Science Basic Research Program of Shaanxi (Grants Nos. 2026JC-YBMS-0012, 2026JC-YBQN-0021).
\appendix

\section{State densities and cooperation preference}\label{sec:Appendix_pdf}

\begin{figure}[htbp]
	\centering
	\includegraphics[width=0.49\textwidth]{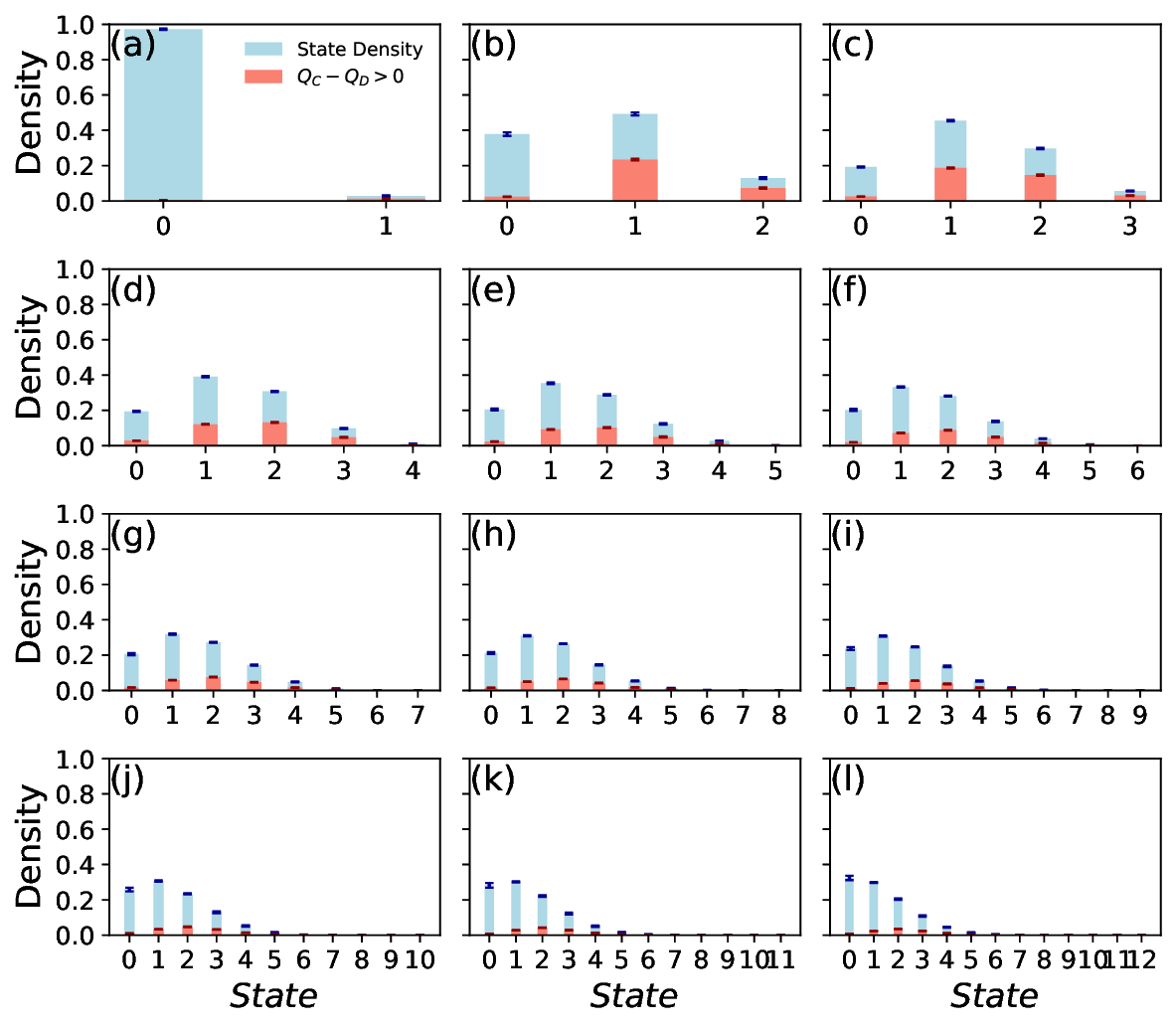}
	\caption{State density and cooperation preference under that density for each state for different neighborhood sizes on the 2d lattice. (a)-(l) corresponds to the state density (in blue) and the fraction of individuals with \({\Delta Q}_{s_j} > 0\) (in orange-red) for each state with neighborhood size \(k = 1\) to \(12\), respectively. The data represent means over the final \(1\times10^{4}\) steps after reaching the steady state, averaged over 20 independent realizations. Error bars represent the standard deviations. Parameters are the same as Fig.~\ref{fig:coop_willingness}.}
	\label{fig:lattice_pdf}
\end{figure}

\begin{figure}[htbp]
	\centering
	\includegraphics[width=0.49\textwidth]{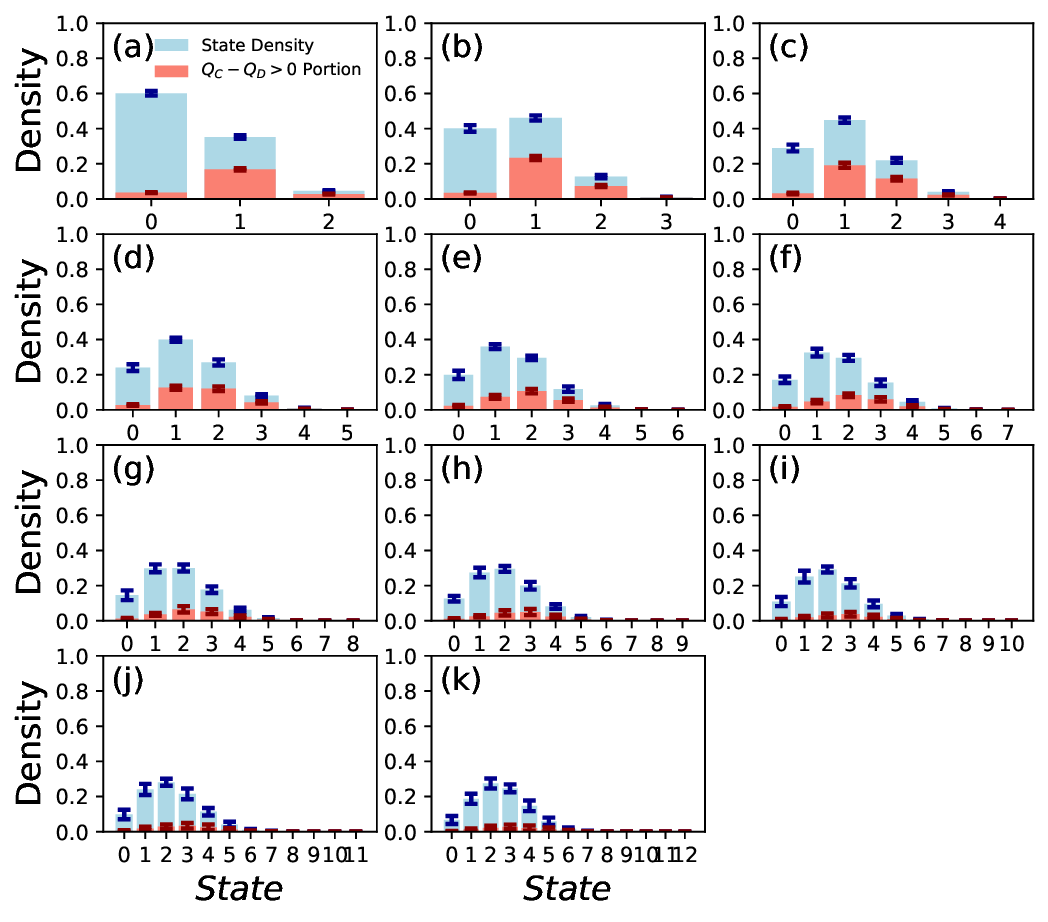}
	\caption{State density and cooperation tendency under that density for each state for different neighborhood sizes on the BA networks. (a)-(k) corresponds to node degree \(k = 2\) to \(12\), respectively. Other setups are the same as Fig.~\ref{fig:lattice_pdf}.}
	\label{fig:BA_pdf}
\end{figure}

To more clearly show the state density and cooperation preference shown in Fig.~\ref{fig:coop_willingness}(a) and~\ref{fig:coop_willingness}(c), here we separately illustrate them in Figs.~\ref{fig:lattice_pdf} and~\ref {fig:BA_pdf}, which present the distributions for 2d square lattice and BA networks, respectively. The state distributions in both figures show that the state densities for large $k$ are vanishing, 

\section{Time evoution of \(\overline{\Delta Q}_{s_j}\) for $k=3$ and 4}\label{sec:Appendix_deltaQ}

\begin{figure}[!htb]
\centering
\includegraphics[width=\columnwidth]{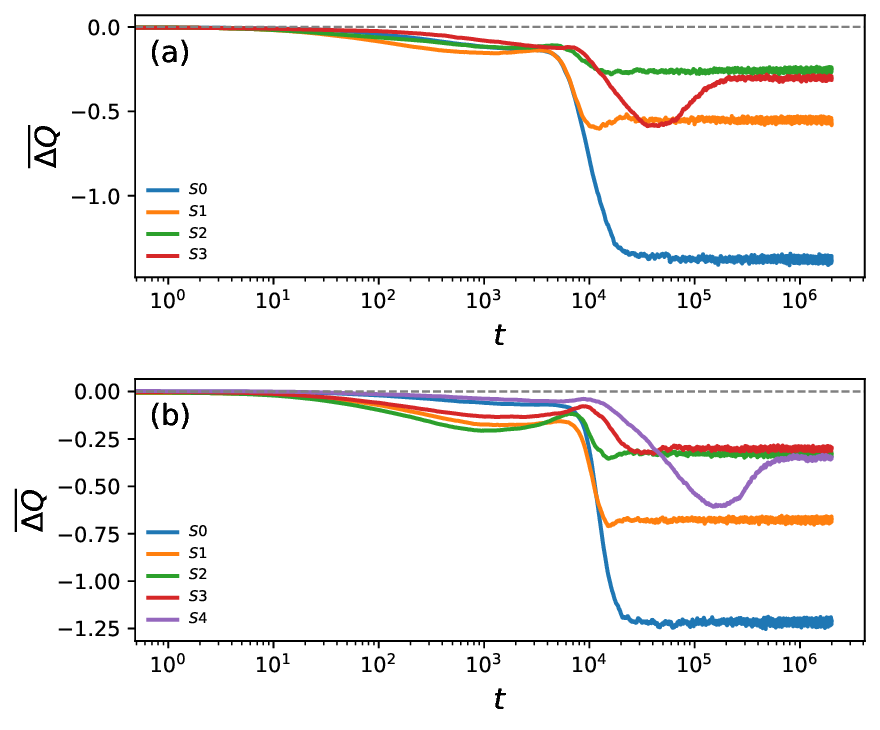}
\caption{Time evolution of \(\overline{\Delta Q}_{s_j}\) for each state in the case of \(k=3\) (a) and \(k=4\) (b) with the ensemble average. 
Each data point is averaged over 50 independent realizations. 
Other parameters: \(t_{\text{max}} = 2\times10^{6}\).}
	\label{fig:deltaQ}
\end{figure}

As shown in Fig.~\ref{fig:deltaQ}, we present the temporal evolution of \(\overline{\Delta Q}_{s_j}\) by ensemble average with the information size that yields a relatively high \(f_C\) on the two-dimensional lattice. It can be observed that \(\overline{\Delta Q}_{s_j} < 0\) for all states, indicating that individuals consistently have a stronger propensity to defect than to cooperate in every state. This explains why, even under the optimal neighborhood size, the overall cooperation level \(f_C\) remains relatively modest. Furthermore, after the system reaches the steady state, notable differences in \(\overline{\Delta Q}_{s_j} < 0\) emerge across states. Specifically, the more cooperators there are among the neighbors, the weaker the individual's tendency to defect -- a pattern that aligns with intuitive expectations.
  
\section{Deviation of the semi-analytic theory}\label{sec:Appendix_theory}

The conditional probability that an individual chooses cooperation within state $s$ is given by
\begin{equation}
	p(C \mid s) = \pi_c(s) \cdot (1 - \varepsilon) + 0.5 \cdot \varepsilon,
	\label{eq:cond_coop_prob}
\end{equation}
where we denote $\pi_c(s)$ as the fraction of $\Delta Q_s>0$ for the whole population, computed from numerical simulations. The second term at the RHS is due to the equal chance of being cooperative in exploration.

The average cooperation level over the entire population is computed as
\begin{equation}
	f_C = \sum_{s=0}^{K} \rho(s) \cdot p(C \mid s),
	\label{eq:global_coop}
\end{equation}
 where $\rho(s)$ denotes the density of state $s$, and $K$ is the neighborhood size we defined in the main text.

Within the mean-field approximation, the cooperation propensity $p(C \mid s_t)$ is assumed to be uniquely determined by the global state $s_t$, which remains identical for the central agent and all its neighbors. Accordingly, the state transition probability is expressed as:
\begin{align}
	P(s_t \rightarrow s_{t+1})
	&= \sum_{a=0}^{K-s_t} \sum_{b=0}^{s_t} \binom{K-s_t}{a} \binom{s_t}{b}
	\bigl[p(C \mid s_t)\bigr]^{a + s_t - b} \notag \\
	&\quad \cdot \bigl[1-p(C \mid s_t)\bigr]^{K - s_t - a + b} \notag \\
	&\quad \cdot \delta\bigl(s_{t+1},\,(s_t - b) + a\bigr),
	\label{eq:state_transition}
\end{align}
where $\delta(\cdot,\cdot)$ is the Kronecker delta function, $a$ is the number of defectors switching to cooperators at a given step, while $b$ denotes the corresponding number of cooperators switching to defectors.
All $P(s_t \rightarrow s_{t+1})$ then constitute the state transition matrix. 
With the state transition matrix, we can compute the state density at time $t+1$:
\begin{equation}
	\rho_{t+1}(s) = \sum_{s_t=0}^{K} P(s_t \rightarrow s) \cdot \rho_t(s_t).
	\label{eq:density_evolution}
\end{equation}
At steady state, the state density is independent of time:
\begin{equation}
	\rho_{t+1}(s) = \rho_t(s), \quad \forall\, s.
	\label{eq:steady_state}
\end{equation}
Combining Eqs.~\eqref{eq:state_transition}--\eqref{eq:steady_state} yields
\begin{equation}
	\rho(s) = \sum_{s_t=0}^{K} P(s_t \rightarrow s) \cdot \rho(s_t).
	\label{eq:density_equation}
\end{equation}
Meanwhile, the stationary density satisfies the normalization constraint:
\begin{equation}
	\sum_{s=0}^{K} \rho(s) = 1.
	\label{eq:normalization}
\end{equation}

Equations \eqref{eq:density_equation} and \eqref{eq:normalization} can be recast into a linear algebraic system
\begin{equation}
	\mathbb{A}\,\mathbf{\rho}=0,
	\label{eq:linear_system}
\end{equation}
\[
\mathbb{A}[s, s_t] =
\begin{cases}
	1 - P(s_t \rightarrow s), & s = s_t, \\[4pt]
	-P(s_t \rightarrow s),     & s \neq s_t.
\end{cases}
\]

Solving the associated eigenvector together with the normalization constraint Eqs.~\eqref{eq:normalization} yields the stationary distribution ${\rho}$.

\end{document}